\documentclass{Interspeech}

\interspeechcameraready

\title{Improving Speech Emotion Recognition Through Cross Modal Attention Alignment and Balanced Stacking Model}

\author[affiliation={1}]{Lucas}{Ueda}
\author[affiliation={1}]{João}{Lima}
\author[affiliation={1,2}]{Leonardo}{Marques}
\author[affiliation={1}]{Paula}{Costa}

\affiliation{}{Universidade Estadual de Campinas (UNICAMP)}{Brazil}
\affiliation{}{CPQD}{Brazil}
  
\email{paulad@unicamp.br}

\keywords{speech emotion recognition, cross-modality, stacking}

\usepackage{comment}
\usepackage{array, booktabs}  
\usepackage{float}
\restylefloat{table}
\usepackage{stfloats}
\usepackage{afterpage}
\usepackage{graphicx}
\begin{document}

\maketitle

\begin{abstract}

Emotion plays a fundamental role in human interaction, and therefore systems capable of identifying emotions in speech are crucial in the context of human-computer interaction. Speech emotion recognition (SER) is a challenging problem, particularly in natural speech and when the available data is imbalanced across emotions. This paper presents our proposed system in the context of the 2025 Speech Emotion Recognition in Naturalistic Conditions Challenge. Our proposed architecture leverages cross-modality, utilizing cross-modal attention to fuse representations from different modalities. To address class imbalance, we employed two training designs: (i) weighted cross-entropy loss (WCE); and (ii) WCE with an additional neutral-expressive soft margin loss and balancing. We trained a total of 12 multimodal models, which were ensembled using a balanced stacking model. Our proposed system achieves a Macro-F1 score of 0.4094 and an accuracy of 0.4128 on 8-class speech emotion recognition.

\end{abstract}

\section{Introduction}






    Conveying emotional content plays a pivotal role in human spoken communication, which makes it essential to consider emotional aspects when developing speech-understanding systems for specific applications. In this scenario, speech emotion recognition (SER) is characterized as the task of automatically identifying a person's emotional state based on their voice \cite{gomez-zaragozaSpeechEmotionRecognition2024}, generally relying on three main steps: data processing, feature selection/extraction, and classification \cite{madanianSpeechEmotionRecognition2023}.

    In healthcare, SER has been explored as a tool to identify emotions expressed by patients with neurological disorders \cite{zisadSpeechEmotionRecognition2020} or to assist clinicians in analyzing depression \cite{zhaoAutomaticAssessmentDepression2020}. In the business sector, successful identification of the emotional state of customers, for example, during call center interactions, can be a way to improve client retention and better assess the quality of service \cite{mekruksavanichNegativeEmotionRecognition2020, vaudableNegativeEmotionsDetection2012}. Furthermore, in-car voice systems with SER capabilities may help prevent dangerous driving situations \cite{jonesUsingParalinguisticCues2008}. Other promising fields of application include human-robot interaction \cite{jonesAffectiveHumanRoboticInteraction2008} and affective computer gaming \cite{jonesAcousticEmotionRecognition2008}, highlighting the versatility of SER technologies.
    
    Although relevant progress has been made in this field, SER still poses a significant challenge even to state-of-the-art speech models \cite{madanianSpeechEmotionRecognition2023}, which highlights the complexity of the task itself. The speech signal is inherently variable and carries a high informational load to account for linguistic and paralinguistic factors, thus rendering classification models great difficulties when trying to identify emotion-related content specifically. A common approach to tackle this problem is hand-engineering sets of acoustic features capable of explaining emotional variance in speech, then using them as input to train the classification system \cite{schullerSpeechEmotionRecognition2004}. More recently, self-supervised deep learning models have shown that it is possible to create meaningful speech representations from raw waveforms \cite{chenWavLMLargeScaleSelfSupervised2022a, hsuHuBERTSelfSupervisedSpeech2021b}. However, these models heavily rely on enormous amounts of training data, a condition not easily met in the context of SER since acquiring emotional speech datasets is challenging by itself. Datasets based on acted speech allow better control over desired emotional performances but may diverge from real-world expressiveness. Acquiring genuinely emotional spontaneous speech, on the other hand, raises questions regarding data privacy and requires human annotation, which is both costly and prone to inter-annotator disagreement \cite{wangMultimodalEmotionRecognition2011}. 

    In this paper, we present a categorical SER system that uses a Random Forest stacking model trained on top of 12 pretrained cross-modal models on the MSP podcast dataset provided by the 2025 Speech Emotion Recognition in Naturalistic Conditions Challenge. The intuition behind using a model ensemble is to combine various data representations that together might be capable of capturing emotional content. Since previous works have shown that textual content representations are capable of enhancing emotion recognition \cite{chen1stPlaceSolution2024}, all models that compose our ensemble are bimodal, combining audio and text features. Additionally, two models also work with prosodic features extracted with NaturalSpeech3's FACodec \cite{juNaturalSpeech3ZeroShot2024} (NS3). A major issue with the dataset we work with is class imbalance, which we tackle using a class-weighted cross-entropy loss function, soft margin loss, and a balanced batching data loading scheme.

\section{Related Works}

Several works have verified that the combination of multiple modalities led to an overall improvement in the general emotion recognition task~\cite{castellano2008emotion}. This is due to the fact that introducing more information helps the model to get a fuller grasp of one's emotional state~\cite{krishna2020multimodal}. Some examples of modalities used to detect emotions in literature are: visual (facial expressions, body gestures, eye gaze, etc); auditory (speech, environmental sounds, etc); physiological (ECG, EEG, etc); and textual~\cite{hazmoune2024using}. 

In the case of SER, numerous studies have demonstrated that combining speech features with their corresponding transcripts can enhance the overall recognition task ~\cite{chen1stPlaceSolution2024, cho2019deep}. Notably, instead of learning features from scratch, these works have also leveraged pre-trained self-supervised representations (SSL) to benefit from large-scale training and learn smaller emotion recognition heads on top. For instance, a combination of Wav2Vec2, RoBERTa, and even Fabnet, a self-supervised framework for learning facial attributes embeddings from video, has been employed to perform SER ~\cite{siriwardhana2020multimodal}.

Moreover, prosody representations obtained by training an LSTM on the GEMAPS~\cite{eyben_gemaps_2016} acoustic parameters, combined with textual features obtained by training a CNN on the speech transcripts, was shown to improve emotion recognition performance by over relative 20\% in terms of unweighted accuracy on the IEMOCAP dataset when compared to solely using audio by and over 9\% over only text~\cite{cho2019deep}. The authors also experiment using both features together with handcrafted pre-trained e-vector features, which represent prosody. The improvement compared to the single audio baseline raised was up to over 21\%. This indicates that scaling up the number of modalities/features, as long as they are diverse in the sense of introducing more information, can improve model performance.




\section{System Description}
Based on the idea that scaling the number of cross-modal features can improve the emotion recognition task, our system consists on a meta-model that is trained on the representations produced by several cross-modal models. The lasts are obtained using a number of combinations of pre-trained features from different modalities and trained independently with a carefully designed features fusion mechanism using cross-modal attention. Our system also employs different losses and batch-balancing strategies to handle class imbalance\footnote{Code is available at: \url{https://github.com/AI-Unicamp/interspeech_ser}}.

\subsection{Cross-Modal Architecture}

\begin{figure}[h]
    \centering
    \includegraphics[width=0.9\linewidth]{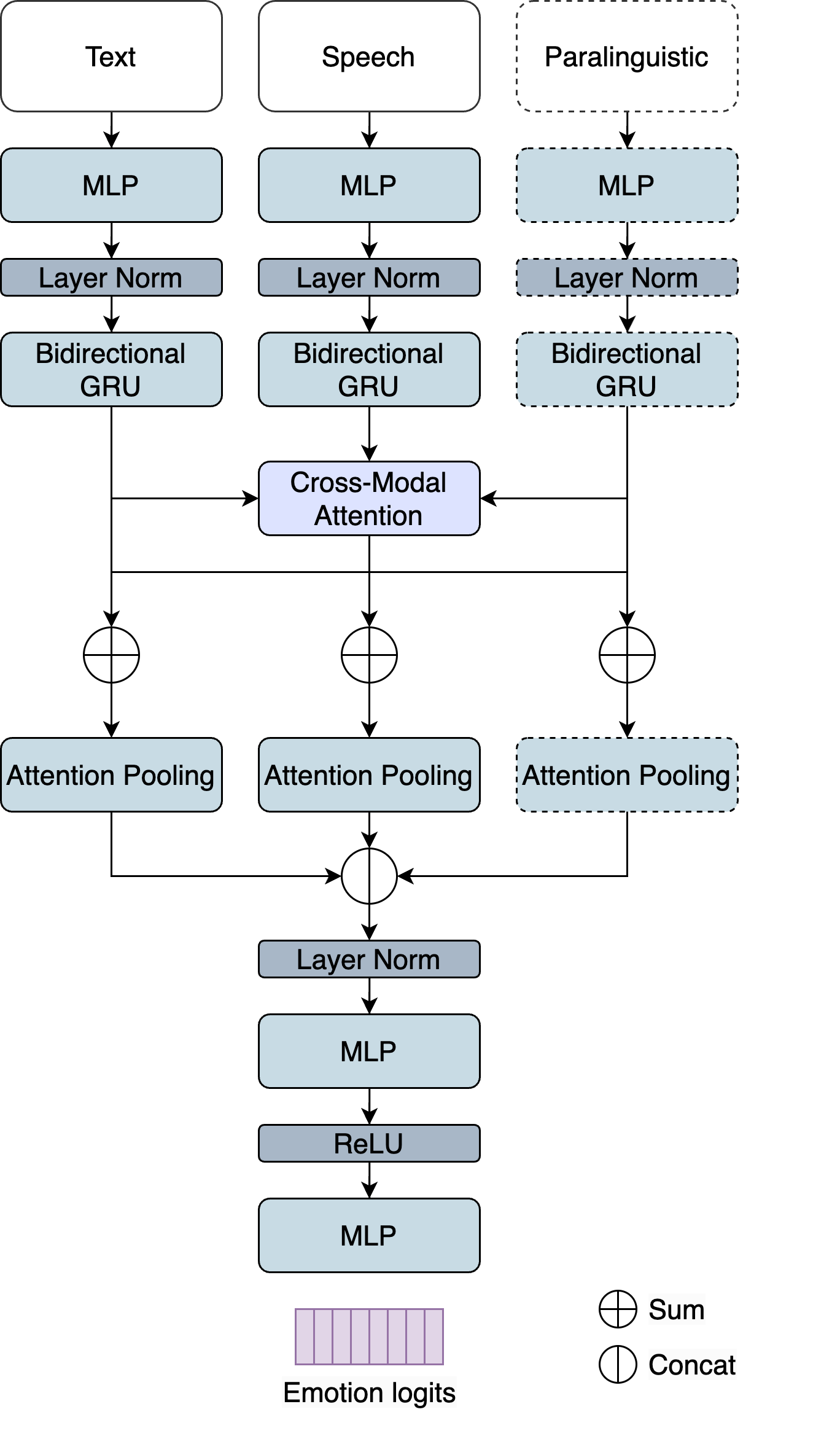}
    \caption{Proposed cross-modal architecture. Dashed lines indicates the tri-modal branch approach when applicable. Each text, speech and paralinguistic inputs are frozen during proposed model training.}
    \label{fig:proposedarq}
\end{figure}

The proposed architecture is illustrated in Figure~\ref{fig:proposedarq}. Text features are encoded at token level $T^{L \times D_t} \in R^{L \times D_t}$ (where $L$ is the number of text tokens and $D_t$ is the dimension of text features), while paralinguistic and speech features are extracted at frame level ($P^{F_p \times D_p} \in R^{F_p \times D_p}$ and $S^{F_s \times D_s} \in R^{F_s \times D_s}$ with $F_p$ and $F_s$ being the paralinguistic and speech number of frames and $D_p$ and $D_s$ the respectively features dimensions). Then both are passed through a Multilayer Perceptron (MLP), projecting them onto a common hidden size $h$ = 512. To normalize the values across all modalities, a layer normalization is applied. Each modality then undergoes a bidirectional Gated Recurrent Unit (GRU), generating a final representation by concatenating the latent layers in each direction ($T^{T \times D_t} \xrightarrow{} T^{T \times D_{2h}} $, $P^{F_p \times D_p} \xrightarrow{} P^{F_p \times D_{2h}} $, $S^{F_s \times D_s} \xrightarrow{} S^{F_s \times D_{2h}} $).

Finally, the modalities are aligned using a cross-modal attention layer. This layer consists of a multi-head attention mechanism that takes one modality as the query and another modality as the key and value, thereby aligning their representations with respect to each other for each distinct modality combination. The output of the cross-modal attention is added to the output of the GRU, balancing the modality-specific information and the cross-modal alignment. This operation is performed for all combinations of modalities present in the architecture. The number of heads is set to 1 to avoid overfitting. After this alignment, the features are aggregated using attention pooling, collapsing each modality into a single dense representation weighted by the learned attention weights ($T^{L \times D_{2h}} \xrightarrow{} T^{1 \times D_{2h}} $, $P^{F_p \times D_{2h}} \xrightarrow{} P^{1 \times D_{2h}} $, $S^{F_p \times D_{2h}} \xrightarrow{} S^{1 \times D_{2h}} $). Equations~\ref{eq:softmax} and~\ref{eq:attention_pool} define the attention pooling operation:

\begin{equation}
    \text{Attention weights $w$}: w_{i} = \frac{\exp(\frac{r_im}{\sqrt{D}})}{\sum_{l=1}^{L} \exp(\frac{r_lL}{\sqrt{D}})}
    \label{eq:softmax}
\end{equation}

\begin{equation}
    \text{Pooled representation } r : r = \sum_{i=1}^{L} w_i r_i  
    \label{eq:attention_pool}
\end{equation}

Where $w$ is the learned attention weights, $r_i \in R^D$ is the modality feature frame/token $i$ with dimension $D$ and $m \in R^D$ is a trainable parameter. The pooled representation $r$ is the weighted average of the modality feature. All pooled representation are concatenated generating a single dense vector $C \in R^{D_f}$, where $D_f$ is the concatenated dimension of all features ($4 \times h$ for bimodal models and $6 \times h$ for trimodal ones). 

Emotion logits are generated by processing $C$ through a classifier layer, which consists of a layer normalization, an MLP, a Rectified Linear Unit (RELU) activation function, and a final MLP that outputs the logits. This classifier layer transforms the dense representation into a set of emotion-specific logits.

\subsubsection{Training design}

To address the imbalance between different emotions, we propose training the model in two different designs. The first consists of using weighted cross-entropy (WCE), where the loss is weighted by the inverse of the frequency of each class, similar to the approach proposed in \cite{chen1stPlaceSolution2024}. We define the weights for each class using the Equation~\ref{eq:wprior}:

\begin{equation}
    w_{prior} = \{w_j = \frac{N}{N_j}, \text{ for } j \in [1,..,E]\}
    \label{eq:wprior}
\end{equation}

where $E$ is the number of different emotions, $N_j$ is the amount of samples of emotion class $j$ in the training set and $N$ is the total amount of samples in training set. This approach allows us to assign higher weights to underrepresented classes and lower weights to overrepresented classes, thereby mitigating the effects of class imbalance and promoting a more balanced training process.

While different emotion classes are imbalanced with respect to each other, partitioning the dataset into expressive speech (all non-neutral labels) and neutral speech (only neutral label) results in a dataset that suffers significantly less from imbalance. In our second training design, we leverage this segmentation to balance the training process. In addition to the weighted cross-entropy (WCE) loss function on emotion labels, we also utilize a Soft Margin Loss (SML) on neutral and expressive samples. SML is responsible to make intermediate layers to model different hidden features for neutral and expressive samples making it easier to further classify the correct emotion label. SML is defined by the Equation~\ref{eq:softmargin}.

\begin{equation}
    \text{SML}(x, y) = \sum_{i=1}^{N} \frac{\log(1 + \exp(-y[i]*x[i]))}{\text{N}}
    \label{eq:softmargin}
\end{equation}

Furthermore, in our batch sampling process, we employ a balancing strategy to ensure that each batch contains a balanced mix of expressive and neutral samples. This approach enables us to increase the variability in the predictive capacity of our cross-modal models.

\subsection{Stacking: Meta Model}

Combining the predictions made by different models is known as ensemble learning. There are several ways to achieve this, with majority voting and logit averaging being the most straightforward approaches. An alternative is the use of stacking ~\cite{wolpert_stacking_1992}, which involves training a meta-model on the logits of different models, leveraging the knowledge extracted by each one from the dataset to train a model that learns how to aggregate this knowledge and improve the final performance. 
By using a meta-model to learn how to weight and combine the predictions of individual models, we can effectively harness the strengths of each model and mitigate their weaknesses, resulting in a more stable and high-performing ensemble.


We propose the use of a 5-fold Random Forest as a meta-model, trained on a balanced training partition where the number of samples for each emotion is defined by the class with the smallest representation. In this way, our meta-model is responsible for aligning the predictions of each cross-modal model, leveraging the different emotions equally. This approach allows the meta-model to effectively combine the strengths of each individual model, while mitigating the effects of class imbalance and promoting a more robust and balanced emotion recognition performance.



\section{Experimental setup}
\subsection{Dataset}

The provided challenge data consist of recordings from the MSP-Podcast dataset~\cite{Naini_2025}. The speaking turns have been perceptually annotated by at least five raters with categorical and attribute-based emotional labels. For categorical classification there are 8 emotional labels: Anger, contempt, Disgust, Fear, Happiness, Neutral, Sadness, and Surprise. The distribution over emotions are imbalanced, ranging from 1.5\% (Fear) to around 40\% (Neutral). The provided train and development (dev) sets are split with an approximate ratio of 73/27\% (66992/25258 total samples, respectively) with no speaker overlaps. A balanced set with 3200 samples is used as test set.

        
        



\subsection{Experiments}

Table~\ref{tab:ssl_specs} provides a comprehensive overview of the Self-Supervised Learning (SSL) models used for each modality in our experiments. The table includes details on the size of each SSL model, the dimensionality of the layer used to extract representations, and the number of hours utilized during pre-training.



        
        
        

\begin{table}[h]

    \caption{Self-supervised models specifications..}
    \label{tab:ssl_specs}
    \centering
     \resizebox{\columnwidth}{!}{ 

    \begin{tabular}{c|c|c|c|c}
        \toprule
        
            Model & Modality & Dim & Hours\\
        \midrule
        
            Whisper Large V3 & Speech & 1280         & 5M\\
            Hubert XL        & Speech & 1280        & 60k\\
            WavLM Large      & Speech & 1024         & 94k\\
            FACodec (NS3)    & Paralinguistic & 512         & 60k\\
            Roberta Large    & Text & 1024         & -\\
            Deberta XXL V2   & Text & 1536         & -\\
        
        \bottomrule
    \end{tabular}
    }
\end{table}

To maintain consistency in training and testing conditions, we generated audio transcriptions using the Whisper Large V3 model, as the transcriptions were not provided for the test set. For speech representation, we utilized Whisper encoder, Hubert, and WavLM models, which demonstrated the best performance in our internal experiments. For textual representation, we employed the Roberta and Deberta models; however, the combination with Deberta was only used with the Whisper speech model, as other combinations did not prove robust enough.

We leveraged the pre-trained FACodec model from Natural Speech 3~\cite{ju2024naturalspeech}, which comprises factorized codecs for text-to-speech synthesis (speech, prosody, text, speaker) to incorporate paralinguistic information. Specifically, we considered two factors: prosody and speaker. We extracted prosodic representations from the FACodec model and concatenated them with speaker representations, resulting in a vector with $L$ frames and a 512 hidden dim. This allowed us to capture nuanced paralinguistic cues, such as tone, pitch, and speaker identity, and integrate them into our model.

All experiments used an initial learning rate of either 1e-4 or 1e-5 chosen according to observed effects and adjusted by a cosine annealing scheduler. Batch size was fixed at 64 and each training lasted between 20 and 30 epochs. All experiments were carried out using the PyTorch deep learning library. The foundational speech models were loaded from HuggingFace's Transformers library and NaturalSpeech 3's FACodec component~\cite{ju2024naturalspeech, zhang2023amphion} was used as provided in the authors' official repository \footnote{\url{https://github.com/lifeiteng/naturalspeech3_facodec/tree/main/ns3_codec}}. Each model that composes our ensemble was trained with a single NVIDIA Quadro RTX 8000 (48GB) GPU and each experiment took around 24 hours for complete training.

Stacking model was based on a 5-fold Random Forest trained on top of a balanced subset of training set. Each RandomForest is defined with 200 estimators, max depth of 8, gini criterion, and both min samples on leaf and min samples split equal to 10.

\subsection{Evaluation}

Macro $F1$ score is used as the official metric. However, dev set still exhibited an imbalance among emotions. To evaluate the model's performance, we employed a bootstrap resampling technique, where we extracted the $F1$ score for 100 balanced subsets of the dev set. We then calculated the mean and minimum-maximum range of the $F1$ values, which we referred to as the $BS-F1$ score and $BS-F1$ range, respectively. This approach allowed us to assess the model's performance in a more robust and reliable manner, taking into account the inherent class imbalance in the data.

\section{Results}

A total of 12 models were trained, comprising 10 bimodal models and 2 trimodal models. Half of these models were trained using only WCE loss, while the remaining half were trained with the additional batch balancing and SML loss. Results are compared with a provided WavLM-based baseline~\cite{goncalves24_odyssey}. The performance metrics, including accuracy, $F1$ score, BS-F1 score, and $BS-F1$ range, are presented in the Table~\ref{tab:results}.

\begin{table*}[ht]
  \caption{Experimental results for all trained models. Balancing stands for whether we use neutral-expressive batch balancing. In the case of Stacking model, balancing refers to the balanced training dataset used in it. Bold values are the best results in each column, while underlined one are the second best.}
  \label{tab:results}
  \centering

  \resizebox{\textwidth}{!}{
  \begin{tabular}{c | c | c | cc | c c | c c}
    \multicolumn{3}{c}{\textbf{}} &  \multicolumn{2}{c}{\textbf{Dev}} &  \multicolumn{2}{c}{\textbf{BS-Dev}} &  \multicolumn{2}{c}{\textbf{Test}}\\ 
    \textbf{Feature combination} & \textbf{Loss function} & \textbf{Balancing} & $F1$ & Acc. & $BS-F1$ & $BS-F1$ Range & $F1$ & Acc. \\
    \midrule

    Baseline & WCE & No & - & - & - & - & 0.329 & 0.356 \\

    Reproduced Baseline & WCE & No & 0.346 & 0.520 & 0.325 & [0.290, 0.356] & 0.284 & 0.320 \\

    \midrule

    WavLM + Roberta & WCE & No & 0.363 & 0.498 & 0.392 & [0.366, 0.423] & - & - \\
    Whisper + Roberta & WCE & No & \textbf{0.388} & 0.524 & \underline{0.417} & \underline{[0.392, 0.440]} & - & - \\
    Whisper + Deberta & WCE & No & 0.376 & 0.527 & 0.383 & [0.364, 0.414]& - & - \\
    Hubert + Roberta & WCE & No & 0.359 & 0.493 & 0.387 & [0.356, 0.418] & - & - \\
    Hubert + Whisper & WCE & No & 0.345 & 0.499 & 0.347 & [0.327, 0.366] & - & - \\
    Whisper + Roberta + NS3 & WCE & No & 0.383 & \textbf{0.547} & 0.374 & [0.347, 0.391] & - & - \\

    WavLM + Roberta & WCE + SML & Yes & 0.375 & 0.505 & 0.398 & [0.364, 0.425] & - & - \\
    Whisper + Roberta & WCE + SML & Yes & \underline{0.387} & 0.529 & 0.410 & [0.388, 0.433] & - & - \\
    Whisper + Deberta & WCE + SML & Yes & \textbf{0.388} & 0.531 & 0.404 & [0.375, 0.431]& - & - \\
    Hubert + Roberta & WCE + SML & Yes & 0.370 & 0.512 & 0.382 & [0.356, 0.408] & - & - \\
    Hubert + Whisper & WCE + SML & Yes & 0.351 & 0.507 & 0.354 & [0.312, 0.411] & - & - \\
    Whisper + Roberta + NS3 & WCE + SML & Yes & 0.386 & \underline{0.542} & 0.391 & [0.338, 0.433] & - & - \\

    \midrule
    Stacking & - & Yes & $0.379$ & $0.498$ & $\textbf{0.430}$ & $[\textbf{0.404},\textbf{0.451}]$ & $\textbf{0.409}$ & $\textbf{0.413}$ \\
    \bottomrule
  \end{tabular}
  }
\end{table*}

Our experimental results demonstrate that the combination of WCE loss, SML loss and batch balancing is effective in most cases, with the balanced models achieving higher $BS-F1$ scores than their unbalanced counterparts. Notable exceptions include the Whisper + Roberta combination with WCE, which also yielded the highest average score with a $BS-F1$ of 0.417, and the Hubert + Roberta model.

We also evaluated two models utilizing different SSL representations from the same modality (speech), specifically the Hubert + Whisper combination. Interestingly, both models exhibited the poorest results in terms of $F1$ and $BS-F1$ scores compared to their bimodal counterparts. This highlights the importance of modality variability in input features for effective classification in the proposed architecture. Nevertheless, both models were considered during the stacking process, as the meta-model was able to extract marginal improvements from their inclusion.

Furthermore, our results show that in terms of accuracy, the best-performing models were the trimodal models, demonstrating the potential of leveraging paralinguistic features when handling naturalistic speech emotion recognition.

Finally, our stacking approach with the meta-model achieved a $BS-F1$ score of 0.430 on the local bootstrap validation, which translated to a score of 0.409 on the test set, outperforming both the reproduced baseline that achieves a test score of 0.284 and the official baseline 0.329. This demonstrates the effectiveness of the proposed model in the categorical task of the challenge, highlighting the benefits of combining multiple models and modalities to improve overall performance.


\section{Conclusion}

Our experiments demonstrate that utilizing SSL representations from different modalities is an effective approach for speech emotion recognition in naturalistic conditions. Notably, the proposed architecture benefits from the combination of multiple modalities, with its worst performance occurring when only a single modality is used. By mixing one or more modalities, the architecture achieves the best results in terms of accuracy or F1 score. The strategies employed to improve the model's performance and maintain its robustness in scenarios with imbalanced classes proved effective. By combining models with WCE loss, SML loss, and balanced batching with a balanced stacking model, we obtained results that outperformed the baseline and remained among the top performers on the challenge leaderboard.

Future work will involve analyzing the use of data augmentation and exploring the emotion adaptation of SSL models before applying the proposed cross-modal architecture. 

\section{Acknowledgements}


This study is partially funded by CAPES – Finance Code 001. It is also supported by FAPESP (BI0S \#2020/09838-0 and Horus \#2023/12865-8). Paula Costa, Lucas Ueda, and João Lima are affiliated with the Dept. of Computer Engineering and Automation (DCA), Faculdade de Engenharia Elétrica e de Computação, and are part of the AI Lab., Recod.ai, Institute of Computing, UNICAMP. Leonardo Marques developed this work while at CPQD and is now at Télécom Paris, Institut Polytechnique de Paris. This project was supported by MCTI under Law 8.248/1991, PPI-Softex, published as Cognitive Architecture (Phase 3), DOU 01245.003479/2024-1.

\bibliographystyle{IEEEtran}
\bibliography{mybib}

\begin{thebibliography}{10}
\providecommand{\url}[1]{#1}
\csname url@samestyle\endcsname
\providecommand{\newblock}{\relax}
\providecommand{\bibinfo}[2]{#2}
\providecommand{\BIBentrySTDinterwordspacing}{\spaceskip=0pt\relax}
\providecommand{\BIBentryALTinterwordstretchfactor}{4}
\providecommand{\BIBentryALTinterwordspacing}{\spaceskip=\fontdimen2\font plus
\BIBentryALTinterwordstretchfactor\fontdimen3\font minus \fontdimen4\font\relax}
\providecommand{\BIBforeignlanguage}[2]{{%
\expandafter\ifx\csname l@#1\endcsname\relax
\typeout{** WARNING: IEEEtran.bst: No hyphenation pattern has been}%
\typeout{** loaded for the language `#1'. Using the pattern for}%
\typeout{** the default language instead.}%
\else
\language=\csname l@#1\endcsname
\fi
#2}}
\providecommand{\BIBdecl}{\relax}
\BIBdecl

\bibitem{gomez-zaragozaSpeechEmotionRecognition2024}
\BIBentryALTinterwordspacing
L.~{G{\'o}mez-Zaragoz{\'a}}, {\'O}.~Valls, R.~del Amor, M.~J. {Castro-Bleda}, V.~Naranjo, M.~A. Raya, and J.~{Mar{\'i}n-Morales}, ``Speech emotion recognition from voice messages recorded in the wild,'' Mar. 2024. [Online]. Available: \url{http://arxiv.org/abs/2403.02167}
\BIBentrySTDinterwordspacing

\bibitem{madanianSpeechEmotionRecognition2023}
S.~Madanian, T.~Chen, O.~Adeleye, J.~M. Templeton, C.~Poellabauer, D.~Parry, and S.~L. Schneider, ``Speech emotion recognition using machine learning --- {{A}} systematic review,'' \emph{Intelligent Systems with Applications}, vol.~20, p. 200266, Nov. 2023.

\bibitem{zisadSpeechEmotionRecognition2020}
S.~N. Zisad, M.~S. Hossain, and K.~Andersson, ``Speech {{Emotion Recognition}} in {{Neurological Disorders Using Convolutional Neural Network}},'' in \emph{Brain {{Informatics}}}, M.~Mahmud, S.~Vassanelli, M.~S. Kaiser, and N.~Zhong, Eds.\hskip 1em plus 0.5em minus 0.4em\relax Cham: Springer International Publishing, 2020, vol. 12241, pp. 287--296.

\bibitem{zhaoAutomaticAssessmentDepression2020}
Z.~Zhao, Z.~Bao, Z.~Zhang, J.~Deng, N.~Cummins, H.~Wang, J.~Tao, and B.~Schuller, ``Automatic {{Assessment}} of {{Depression From Speech}} via a {{Hierarchical Attention Transfer Network}} and {{Attention Autoencoders}},'' \emph{IEEE Journal of Selected Topics in Signal Processing}, vol.~14, no.~2, pp. 423--434, Feb. 2020.

\bibitem{mekruksavanichNegativeEmotionRecognition2020}
S.~Mekruksavanich, A.~Jitpattanakul, and N.~Hnoohom, ``Negative {{Emotion Recognition}} using {{Deep Learning}} for {{Thai Language}},'' in \emph{2020 {{Joint International Conference}} on {{Digital Arts}}, {{Media}} and {{Technology}} with {{ECTI Northern Section Conference}} on {{Electrical}}, {{Electronics}}, {{Computer}} and {{Telecommunications Engineering}} ({{ECTI DAMT}} \& {{NCON}})}, Mar. 2020, pp. 71--74.

\bibitem{vaudableNegativeEmotionsDetection2012}
C.~Vaudable and L.~Devillers, ``Negative emotions detection as an indicator of dialogs quality in call centers,'' in \emph{2012 {{IEEE International Conference}} on {{Acoustics}}, {{Speech}} and {{Signal Processing}} ({{ICASSP}})}, Mar. 2012, pp. 5109--5112.

\bibitem{jonesUsingParalinguisticCues2008}
C.~Jones and I.-M. Jonsson, ``Using {{Paralinguistic Cues}} in {{Speech}} to {{Recognise Emotions}} in {{Older Car Drivers}},'' in \emph{Affect and {{Emotion}} in {{Human-Computer Interaction}}}, C.~Peter and R.~Beale, Eds.\hskip 1em plus 0.5em minus 0.4em\relax Berlin, Heidelberg: Springer Berlin Heidelberg, 2008, vol. 4868, pp. 229--240.

\bibitem{jonesAffectiveHumanRoboticInteraction2008}
C.~Jones and A.~Deeming, ``Affective {{Human-Robotic Interaction}},'' in \emph{Affect and {{Emotion}} in {{Human-Computer Interaction}}}, C.~Peter and R.~Beale, Eds.\hskip 1em plus 0.5em minus 0.4em\relax Berlin, Heidelberg: Springer Berlin Heidelberg, 2008, vol. 4868, pp. 175--185.

\bibitem{jonesAcousticEmotionRecognition2008}
C.~Jones and J.~Sutherland, ``Acoustic {{Emotion Recognition}} for {{Affective Computer Gaming}},'' in \emph{Affect and {{Emotion}} in {{Human-Computer Interaction}}}, C.~Peter and R.~Beale, Eds.\hskip 1em plus 0.5em minus 0.4em\relax Berlin, Heidelberg: Springer Berlin Heidelberg, 2008, vol. 4868, pp. 209--219.

\bibitem{schullerSpeechEmotionRecognition2004}
B.~Schuller, G.~Rigoll, and M.~Lang, ``Speech emotion recognition combining acoustic features and linguistic information in a hybrid support vector machine-belief network architecture,'' in \emph{2004 {{IEEE International Conference}} on {{Acoustics}}, {{Speech}}, and {{Signal Processing}}}, vol.~1, May 2004, pp. I--577.

\bibitem{chenWavLMLargeScaleSelfSupervised2022a}
S.~Chen, C.~Wang, Z.~Chen, Y.~Wu, S.~Liu, Z.~Chen, J.~Li, N.~Kanda, T.~Yoshioka, X.~Xiao, J.~Wu, L.~Zhou, S.~Ren, Y.~Qian, Y.~Qian, J.~Wu, M.~Zeng, X.~Yu, and F.~Wei, ``Wavlm: Large-scale self-supervised pre-training for full stack speech processing,'' \emph{IEEE Journal of Selected Topics in Signal Processing}, vol.~16, no.~6, pp. 1505--1518, 2022.

\bibitem{hsuHuBERTSelfSupervisedSpeech2021b}
\BIBentryALTinterwordspacing
W.-N. Hsu, B.~Bolte, Y.-H.~H. Tsai, K.~Lakhotia, R.~Salakhutdinov, and A.~Mohamed, ``Hubert: Self-supervised speech representation learning by masked prediction of hidden units,'' \emph{IEEE/ACM Trans. Audio, Speech and Lang. Proc.}, vol.~29, p. 3451–3460, Oct. 2021. [Online]. Available: \url{https://doi.org/10.1109/TASLP.2021.3122291}
\BIBentrySTDinterwordspacing

\bibitem{wangMultimodalEmotionRecognition2011}
S.~Haq and P.~Jackson, ``Multimodal {{Emotion Recognition}},'' in \emph{Machine {{Audition}}: {{Principles}}, {{Algorithms}} and {{Systems}}}, W.~Wang, Ed.\hskip 1em plus 0.5em minus 0.4em\relax IGI Global, 2011, pp. 398--423.

\bibitem{chen1stPlaceSolution2024}
M.~Chen, H.~Zhang, Y.~Li, J.~Luo, W.~Wu, Z.~Ma, P.~Bell, C.~Lai, J.~D. Reiss, L.~Wang, P.~C. Woodland, X.~Chen, H.~Phan, and T.~Hain, ``1st {{Place Solution}} to {{Odyssey Emotion Recognition Challenge Task1}}: {{Tackling Class Imbalance Problem}},'' in \emph{The {{Speaker}} and {{Language Recognition Workshop}} ({{Odyssey}} 2024)}.\hskip 1em plus 0.5em minus 0.4em\relax ISCA, Jun. 2024, pp. 260--265.

\bibitem{juNaturalSpeech3ZeroShot2024}
Z.~Ju, Y.~Wang, K.~Shen, X.~Tan, D.~Xin, D.~Yang, Y.~Liu, Y.~Leng, K.~Song, S.~Tang, Z.~Wu, T.~Qin, X.-Y. Li, W.~Ye, S.~Zhang, J.~Bian, L.~He, J.~Li, and S.~Zhao, ``Naturalspeech 3: zero-shot speech synthesis with factorized codec and diffusion models,'' in \emph{Proceedings of the 41st International Conference on Machine Learning}, ser. ICML'24.\hskip 1em plus 0.5em minus 0.4em\relax JMLR.org, 2024.

\bibitem{castellano2008emotion}
G.~Castellano, L.~Kessous, and G.~Caridakis, ``Emotion recognition through multiple modalities: face, body gesture, speech,'' \emph{Affect and Emotion in Human-Computer Interaction: From Theory to Applications}, pp. 92--103, 2008.

\bibitem{krishna2020multimodal}
D.~Krishna and A.~Patil, ``Multimodal emotion recognition using cross-modal attention and 1d convolutional neural networks.'' in \emph{Interspeech}, 2020, pp. 4243--4247.

\bibitem{hazmoune2024using}
S.~Hazmoune and F.~Bougamouza, ``Using transformers for multimodal emotion recognition: Taxonomies and state of the art review,'' \emph{Engineering Applications of Artificial Intelligence}, vol. 133, p. 108339, 2024.

\bibitem{cho2019deep}
J.~Cho, R.~Pappagari, P.~Kulkarni, J.~Villalba, Y.~Carmiel, and N.~Dehak, ``Deep neural networks for emotion recognition combining audio and transcripts,'' in \emph{Interspeech 2018}, 2018, pp. 247--251.

\bibitem{siriwardhana2020multimodal}
S.~Siriwardhana, T.~Kaluarachchi, M.~Billinghurst, and S.~Nanayakkara, ``Multimodal emotion recognition with transformer-based self supervised feature fusion,'' \emph{Ieee Access}, vol.~8, pp. 176\,274--176\,285, 2020.

\bibitem{eyben_gemaps_2016}
F.~Eyben, K.~R. Scherer, B.~W. Schuller, J.~Sundberg, E.~André, C.~Busso, L.~Y. Devillers, J.~Epps, P.~Laukka, S.~S. Narayanan, and K.~P. Truong, ``The geneva minimalistic acoustic parameter set (gemaps) for voice research and affective computing,'' \emph{IEEE Transactions on Affective Computing}, vol.~7, no.~2, pp. 190--202, 2016.

\bibitem{wolpert_stacking_1992}
\BIBentryALTinterwordspacing
D.~H. Wolpert, ``Stacked generalization,'' \emph{Neural Networks}, vol.~5, no.~2, pp. 241--259, 1992. [Online]. Available: \url{https://www.sciencedirect.com/science/article/pii/S0893608005800231}
\BIBentrySTDinterwordspacing

\bibitem{Naini_2025}
A.~R. Naini, L.~Goncalves, A.~N. Salman, P.~Mote, I.~R. Ülgen, T.~Thebaud, L.~Velazquez, L.~P. Garcia, N.~Dehak, B.~Sisman, and C.~Busso, ``The interspeech 2025 challenge on speech emotion recognition in naturalistic conditions,'' in \emph{Interspeech 2025}, vol. To appear, Rotterdam, The Netherlands, August 2025.

\bibitem{ju2024naturalspeech}
Z.~Ju, Y.~Wang, K.~Shen, X.~Tan, D.~Xin, D.~Yang, Y.~Liu, Y.~Leng, K.~Song, S.~Tang \emph{et~al.}, ``Naturalspeech 3: Zero-shot speech synthesis with factorized codec and diffusion models,'' \emph{arXiv preprint arXiv:2403.03100}, 2024.

\bibitem{zhang2023amphion}
X.~Zhang, L.~Xue, Y.~Gu, Y.~Wang, H.~He, C.~Wang, X.~Chen, Z.~Fang, H.~Chen, J.~Zhang, T.~Y. Tang, L.~Zou, M.~Wang, J.~Han, K.~Chen, H.~Li, and Z.~Wu, ``Amphion: An open-source audio, music and speech generation toolkit,'' \emph{arXiv}, vol. abs/2312.09911, 2024.

\bibitem{goncalves24_odyssey}
L.~Goncalves, A.~N. Salman, A.~R. Naini, L.~Moro-Velázquez, T.~Thebaud, P.~Garcia, N.~Dehak, B.~Sisman, and C.~Busso, ``Odyssey 2024 - speech emotion recognition challenge: Dataset, baseline framework, and results,'' in \emph{The Speaker and Language Recognition Workshop (Odyssey 2024)}, 2024, pp. 247--254.

\end{thebibliography}

\end{document}